\documentclass[twocolumn,showpacs,aps,prl,superscriptaddress]{revtex4}

\usepackage{graphicx}
\usepackage{dcolumn}
\usepackage{amsmath}
\usepackage{epsfig}

\RequirePackage{xspace}

\usepackage{relsize}
\def\babar{\mbox{\slshape B\kern-0.1em{\smaller A}\kern-0.1em
    B\kern-0.1em{\smaller A\kern-0.2em R}}}

\def\epem       {\ensuremath{e^+e^-}\xspace}

\def\mumu       {\ensuremath{\mu^+\mu^-}\xspace}

\def\ellell     {\ensuremath{\ell^+ \ell^-}\xspace}

\def\W      {\ensuremath{W}\xspace}

\def\qqbar {\ensuremath{q\overline q}\xspace}

\def\piz   {\ensuremath{\pi^0}\xspace}

\def\Kbar  {\kern 0.2em\overline{\kern -0.2em K}{}\xspace}

\def\Kz    {\ensuremath{K^0}\xspace}
\def\Kzb   {\ensuremath{\Kbar^0}\xspace}
\def\KzKzb {\ensuremath{\Kz \kern -0.16em \Kzb}\xspace}
\def\Kp    {\ensuremath{K^+}\xspace}
\def\Km    {\ensuremath{K^-}\xspace}

\def\KpKm  {\ensuremath{\Kp \kern -0.16em \Km}\xspace}
 
\def\KL    {\ensuremath{K^0_{\scriptscriptstyle L}}\xspace} 
\def\Kstarz  {\ensuremath{K^{*0}}\xspace}

\def\Dbar    {\kern 0.2em\overline{\kern -0.2em D}{}\xspace}

\def\Dz      {\ensuremath{D^0}\xspace}
\def\Dzb     {\ensuremath{\Dbar^0}\xspace}
\def\DzDzb   {\ensuremath{\Dz {\kern -0.16em \Dzb}}\xspace}
\def\Dp      {\ensuremath{D^+}\xspace}
\def\Dm      {\ensuremath{D^-}\xspace}

\def\DpDm    {\ensuremath{\Dp {\kern -0.16em \Dm}}\xspace}

\def\B       {\ensuremath{B}\xspace}
\def\Bbar    {\kern 0.18em\overline{\kern -0.18em B}{}\xspace}

\def\BB      {\ensuremath{B\Bbar}\xspace} 
\def\Bz      {\ensuremath{B^0}\xspace}
\def\Bzb     {\ensuremath{\Bbar^0}\xspace}
\def\BzBzb   {\ensuremath{\Bz {\kern -0.16em \Bzb}}\xspace}
\def\Bu      {\ensuremath{B^+}\xspace}
\def\Bub     {\ensuremath{B^-}\xspace}

\def\BpBm    {\ensuremath{\Bu {\kern -0.16em \Bub}}\xspace}

\def\jpsi     {\ensuremath{{J\mskip -3mu/\mskip -2mu\psi\mskip 2mu}}\xspace}

\mathchardef\Upsilon="7107
\def\Y#1S{\ensuremath{\Upsilon{(#1S)}}\xspace}% no space before {...}!

\def\FourS {\Y4S}

\mathchardef\Deltares="7101
\mathchardef\Xi="7104
\mathchardef\Lambda="7103
\mathchardef\Sigma="7106
\mathchardef\Omega="710A

\def\Deltabar{\kern 0.25em\overline{\kern -0.25em \Deltares}{}\xspace}
\def\Lbar{\kern 0.2em\overline{\kern -0.2em\Lambda\kern 0.05em}\kern-0.05em{}\xspace}
\def\Sigbar{\kern 0.2em\overline{\kern -0.2em \Sigma}{}\xspace}
\def\Xibar{\kern 0.2em\overline{\kern -0.2em \Xi}{}\xspace}
\def\Obar{\kern 0.2em\overline{\kern -0.2em \Omega}{}\xspace}
\def\Nbar{\kern 0.2em\overline{\kern -0.2em N}{}\xspace}
\def\Xb{\kern 0.2em\overline{\kern -0.2em X}{}\xspace}

\def\BR         {{\ensuremath{\cal B}\xspace}}

\def\upsbb   {\ensuremath{\FourS \to \BB}\xspace}

\def\mes        {\mbox{$m_{\rm ES}$}\xspace}

\def\DeltaE     {\mbox{$\Delta E$}\xspace}

\newcommand{\tev}{\ensuremath{\mathrm{\,Te\kern -0.1em V}}\xspace}
\newcommand{\gev}{\ensuremath{\mathrm{\,Ge\kern -0.1em V}}\xspace}
\newcommand{\mev}{\ensuremath{\mathrm{\,Me\kern -0.1em V}}\xspace}
\newcommand{\kev}{\ensuremath{\mathrm{\,ke\kern -0.1em V}}\xspace}
\newcommand{\ev}{\ensuremath{\mathrm{\,e\kern -0.1em V}}\xspace}
\newcommand{\gevc}{\ensuremath{{\mathrm{\,Ge\kern -0.1em V\!/}c}}\xspace}
\newcommand{\mevc}{\ensuremath{{\mathrm{\,Me\kern -0.1em V\!/}c}}\xspace}
\newcommand{\gevcc}{\ensuremath{{\mathrm{\,Ge\kern -0.1em V\!/}c^2}}\xspace}
\newcommand{\mevcc}{\ensuremath{{\mathrm{\,Me\kern -0.1em V\!/}c^2}}\xspace}
\def\cm   {\ensuremath{\rm \,cm}\xspace}

\def\mum  {\ensuremath{\,\mu\rm m}\xspace}%% mu meter 
\def\invfb   {\ensuremath{\mbox{\,fb}^{-1}}\xspace}

\def\mus  {\ensuremath{\rm \,\mus}\xspace}

\def\mus        {\ensuremath{\,\mu{\rm s}}\xspace}    %% microsecond
\def\rad{\ensuremath{\rm \,rad}\xspace}

\def\to                 {\ensuremath{\rightarrow}\xspace}

\def\pep2{PEP-II}

\newcommand{\dedx}{\ensuremath{\mathrm{d}\hspace{-0.1em}E/\mathrm{d}x}\xspace}

\def\gsim{{~\raise.15em\hbox{$>$}\kern-.85em
          \lower.35em\hbox{$\sim$}~}\xspace}
\def\lsim{{~\raise.15em\hbox{$<$}\kern-.85em
          \lower.35em\hbox{$\sim$}~}\xspace}

\xspace

\newcommand{\epjBase}        {Eur.\ Phys.\ Jour.\xspace}
\newcommand{\jprlBase}       {Phys.\ Rev.\ Lett.\xspace}
\newcommand{\jprBase}        {Phys.\ Rev.\xspace}
\newcommand{\jplBase}        {Phys.\ Lett.\xspace}

\newcommand{\nimBaseC}       {Nucl.\ Instr.\ and Methods\xspace}

\newcommand{\epjc}      [1]  {\epjBase\ C~{\bf #1}}

\newcommand{\nim}       [1]  {\nimBaseC~{\bf #1}}
\newcommand{\plb}       [1]  {\jplBase\ B~{\bf #1}}

\newcommand{\jprl}      [1]  {\jprlBase\ {\bf #1}}
\newcommand{\jprd}      [1]  {\jprBase\ D~{\bf #1}}
\def\jetset74   {\mbox{\tt Jetset \hspace{-0.5em}7.\hspace{-0.2em}4}\xspace}

\def\jpsill {\ensuremath{{\jpsi\to\ellell}}}
\def\jpsiee {\ensuremath{{\jpsi\to\epem}}}
\def\jpsimumu {\ensuremath{{\jpsi\to\mumu}}}
\def\mydecay {\ensuremath{\Bz\to\jpsi\gamma}}

\def\pigg{\ensuremath{\piz\to\gamma\gamma}}
\def\myupsbzbz{\ensuremath{\FourS \to \BzBzb}}

\def\mjeelow{\ensuremath{3.06}}
\def\mjeehigh{\ensuremath{3.12}}
\def\mjmumulow{\ensuremath{3.07}}
\def\mjmumuhigh{\ensuremath{3.13}}
\def\LAThigh{\ensuremath{0.35}}
\def\costhetagamlow{\ensuremath{-0.35}}
\def\costbbzhigh{\ensuremath{0.75}}
\def\costhsphbzhigh{\ensuremath{0.85}}
\def\R2allhigh{\ensuremath{0.45}}
\def\coshelbzhigh{\ensuremath{0.90}}

\def\lowAWmes{\ensuremath{5.2}}
\def\highAWmes{\ensuremath{5.3}}

\def\highAWdele{\ensuremath{0.30}}

\def\LSBmes{\ensuremath{5.270}}
\def\HSBmes{\ensuremath{5.290}}
\def\LSBdele{\ensuremath{-0.05}}
\def\HSBdele{\ensuremath{0.08}}

\def\newuplim{\ensuremath{1.6 \times 10^{-6}}}

\def\costhetathrust{\ensuremath{\cos\theta_{t}}}
\def\costhetasph{\ensuremath{\cos\theta_{sph}}}
\def\costhetaB{\ensuremath{\cos\theta_{B}}}

\def\eff{\ensuremath{\varepsilon_{s}}}

\newcommand{\BABARPubYear}    {04}
\newcommand{\BABARPubNumber}  {23}

\newcommand{\SLACPubNumber} {10585}

\def\figurebox#1#2#3{%
    \def\arg{#3}%
    \ifx\arg\empty
    {\hfill\vbox{\hsize#2\hrule\hbox to #2{\vrule\hfill\vbox to #1{\hsize#2\vfill}\vrule}\hrule}\hfill}%
    \else
    {\hfill\epsfbox{#3}\hfill}%
    \fi}

\begin{document}

\preprint{\babar-PUB-\BABARPubYear/\BABARPubNumber} 
\preprint{SLAC-PUB-\SLACPubNumber} 

\begin{flushleft}
\babar-PUB-\BABARPubYear/\BABARPubNumber\\
SLAC-PUB-\SLACPubNumber\\
%hep-ex/\LANLNumber\\[10mm]
\end{flushleft}

\title{\begin{flushleft}
%\mbox{\normalsize {\babar\ }Analysis Document \#910 Version 9}
\end{flushleft}
\vskip 20pt
{\large \bf
Search for the decay \mydecay} 
}

%\title{
%{\large \bf
%Search for the decay \mydecay} 
%}

% Dummy author list; contact PubBoard Chair for current author list
%\input authors.tex
%% author list as of 02-Jun-2004 (610 authors)
%
\author{B.~Aubert}
\author{R.~Barate}
\author{D.~Boutigny}
\author{F.~Couderc}
\author{J.-M.~Gaillard}
\author{A.~Hicheur}
\author{Y.~Karyotakis}
\author{J.~P.~Lees}
\author{V.~Tisserand}
\author{A.~Zghiche}
\affiliation{Laboratoire de Physique des Particules, F-74941 Annecy-le-Vieux, France }
% ...
\collaboration{The \babar\ Collaboration}
\noaffiliation

\date{\today}% It is always \today, today, but you may specify any date with \date.

\begin{abstract}
We present the results of a search for the radiative decay \mydecay\ in a data set containing 123 million \upsbb\ decays, collected by the \babar\ detector at the PEP-II asymmetric-energy \epem\ storage ring at SLAC.  We find no evidence for a signal and place an upper limit of $\BR(\mydecay) < \newuplim$ at 90\% confidence level.
\end{abstract}

% PACS, the Physics and Astronomy Classification Scheme.

\maketitle

\begin{figure}
\includegraphics[width=0.5\linewidth]{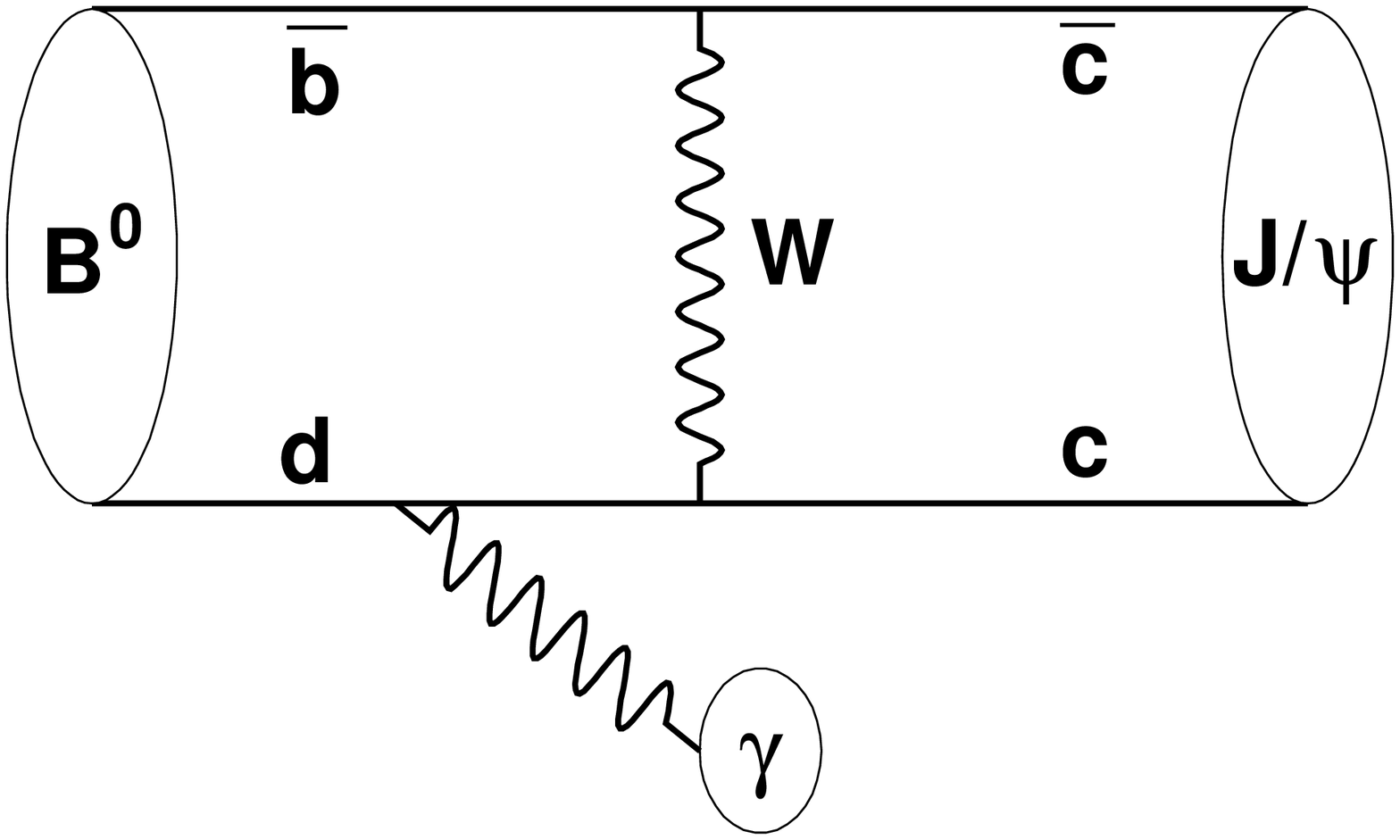}
\caption{Feynman diagram of the leading-order contribution to \mydecay.}\label{fig:leading}
\end{figure}

Rare decays are sensitive probes of possible new physics effects beyond the Standard Model.  The decay $\mydecay^1$ is a very rare decay, with a predicted branching fraction of $7.65 \times 10^{-9}$ \cite{ref:lu}.  The dominant mechanism is the exchange of a \W-boson and the radiation of a photon from the light quark of the \B\ meson (Fig.~\ref{fig:leading}).  Possible new physics enhancements of the \mydecay\ decay rate include a right-handed charged current \cite{ref:lu} or non-spectator intrinsic charm in the \Bz\ meson \cite{ref:gardner}.  No prior search has been conducted for this decay mode.
\footnotetext[1]{Charge conjugation is implied throughout this paper.}

The data used in this analysis were collected with the \babar\ detector at the \pep2\ asymmetric-energy \epem\ storage ring.  The data sample contains a time-integrated luminosity of 113\invfb of ``on-peak'' data taken at the $\Upsilon(4S)$ resonance, corresponding to a center-of-mass energy of $\sqrt{s}= 10.58\gev$, as well as 12\invfb of ``off-peak'' data recorded at about 40\mev\ below this energy.  The energy asymmetry between the low-energy positron beam (3.1\gev) and the high-energy electron beam (9.0\gev) produces a Lorentz boost of $\beta = 0.56$ of the center-of-mass frame with respect to the laboratory frame.  In this report, measurements are presented in the laboratory frame unless otherwise specified.  The $z$-axis points along the direction of the high-energy electron beam; polar angles ($\theta$) are measured with respect to this axis.

The \babar\ detector is described elsewhere~\cite{ref:babar}; here we provide a brief overview.  The detector consists of five subdetectors.  The tracking system includes a 40-layer, helium-based drift chamber (DCH) as the main tracking chamber, and a five-layer silicon vertex tracker (SVT) for precise reconstruction of track angles and \B-decay vertices.  The tracking system covers the polar angular region $0.41 < \theta < 2.54$ \rad (86\% of the solid angle in the center-of-mass frame).  Charged-particle identification, particularly $K/\pi$ separation, is provided by the DIRC, a ring-imaging Cherenkov detector.  Electrons and photons with energy greater than 30\mev and a polar angle within $0.41 < \theta < 2.41$\ rad (84\% of the solid angle in the center-of-mass frame) are detected by an electromagnetic calorimeter (EMC) with energy resolution at 1\gev\ of 2.6\%.  These four subdetectors are contained inside a magnetic solenoid, which supplies a 1.5-T magnetic field for tracking.  The fifth subdetector, a segmented iron flux return (the IFR), surrounds the magnetic solenoid and is instrumented with resistive plate chambers for muon and \KL\ identification.

Particle candidates are either charged tracks in the tracking devices (SVT, DCH), or ``clusters''---groups of adjacent hits---in the EMC or the IFR.  Each charged track is tested to see if it comes from the same particle as one of the clusters, and if so, it is matched with that cluster.  Charged-particle candidates are thus either stand-alone charged tracks or track-cluster pairs, while neutral-particle candidates are clusters not matched with charged tracks.

The analysis proceeds as follows.  We use simulated signal and background samples to derive an optimized set of selection criteria, and to estimate the fractions of signal and background events that pass the criteria.  We then apply the selection criteria to the data sample and calculate an upper limit on the branching fraction for \mydecay. The simulated signal sample contains only \mydecay\ events in which the \jpsi\ meson decays in the \jpsill\ mode, where $\ell$ denotes an electron or a muon.  The sources of background in this analysis include background from \B\ decays and from continuum quark production (\epem\to\qqbar, where $q$ = $u$, $d$, $s$, $c$), the latter being three times the size of the former.  The simulated background sample contains both of these types of background.

To obtain a \BB-enriched sample, we impose requirements optimized independently of this analysis and used in many other \B-decay studies.  Events are required to have visible energy greater than 4.5\gev\ and a ratio of the second to the zeroth Fox-Wolfram moment~\cite{ref:fox}, $R_2$, less than 0.5.  We reconstruct a primary event vertex from the charged tracks and require that it be located within 6\cm\ of the beam spot in the direction parallel to the beam line, and within a transverse distance of 0.5\cm\ from the beam line. The beam spot RMS size is approximately 0.9\cm\ in $z$, 120\mum\ horizontally, and 5.6\mum\ vertically.  There must be at least three tracks in the fiducial volume satisfying the following criteria: they must have transverse momentum greater than 0.1\gevc, momentum smaller than 10\gevc, and at least 12 hits in the DCH; and they must approach within 10\cm\ of the beam spot in $z$ and within 1.5\cm\ of the beam line.  Studies with simulated samples indicate that these criteria are satisfied by 96\% of \upsbb\ decays.

Candidate \mydecay\ decays are reconstructed as follows.  A \Bz\ candidate is formed from a \jpsi\ and a photon candidate.  The \jpsi\ candidate is reconstructed in the low-background, high-efficiency \jpsill\ mode only.  Electron candidates are identified using the ratio of calorimeter energy to track momentum ($E/p$), the ionization loss in the tracking system (\dedx), and the shape of the shower in the calorimeter.  Whenever possible, photons radiated by an electron traversing material prior to the DCH (0.04 radiation lengths at normal incidence) are combined with the track.  These bremsstrahlung-photon candidates are characterized by an EMC energy greater than 30\mev and a polar angle within 35 mrad of the electron direction, as well as an azimuthal angle either within 50 mrad of the electron direction, or between the electron direction at the origin and the azimuth of the impact point in the EMC.  Muons are identified by the energy deposited in the EMC, the compatibility of the track formed by the hits in the IFR with the extrapolation of a track measured in the DCH, and the amount of iron penetrated by this track.  Studies of data-derived control samples show that at a typical lepton momentum of 2\gevc, the efficiency of the electron (muon) identification criteria is 93\% (83\%), with a pion misidentification probability of 0.2\% (8\%).  Photons are neutral candidates with characteristic electromagnetic shower shapes in the EMC.  To determine the photon direction, we assume that the photon candidate originates at the \jpsill\ vertex.

We use the simulated samples to derive an optimized set of selection criteria for \mydecay\ events.  For the optimization we minimize the ratio $\sqrt{\varepsilon_{b}}/\eff$, where $\varepsilon_{b}$ and \eff\ are the respective efficiencies for background and signal events to pass the selection criteria.  The optimized selection criteria are described below and summarized in Table~\ref{tab:selection}.
  
\begin{figure} %[!htb]
\begin{center}
\includegraphics[width=0.88\linewidth]{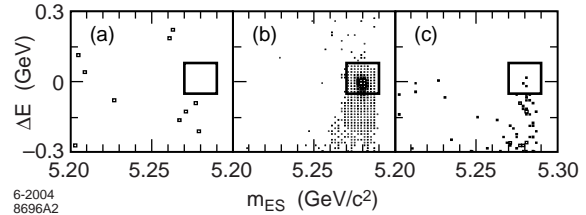}
\caption{\DeltaE-\mes\ distributions in the analysis window for (a) on-peak data, (b) simulated signal, and (c) simulated background.  The signal region is indicated by a box.  The sample in (c) is the original background sample used to optimize the selection; the equivalent luminosity of this sample is about nine times that of the 113\invfb\ sample in (a).}
\label{fig:result}
\end{center}
\end{figure}

To identify and select \B\ candidates we use the kinematic variables \DeltaE\ and \mes.  The energy difference \DeltaE is given by \DeltaE = $(2q_\Upsilon \cdot q_B - s)/2\sqrt{s}$, where $q_\Upsilon = (E_{\Upsilon}, \vec{p}_{\Upsilon})$ is the four-momentum of the \FourS as determined from beam parameters, $q_B = q_{\jpsi} + q_{\gamma} = (E_B, \vec{p}_{B})$ is the reconstructed four-momentum of the \B\ candidate, and $s \equiv q_{\Upsilon}^2$ is the squared center-of-mass energy.  The energy-substituted mass \mes\ is given by $\mes = \sqrt{(s/2 + \vec{p}_{\Upsilon} \cdot \vec{p}_B)^2 / E_\Upsilon^2 - |\vec{p_B}|^2}$.  The advantage of using \DeltaE\ and \mes\ to impose the kinematic constraints for \B\ decays is that these quantities are largely uncorrelated and make maximum use of the well-determined beam four-momentum.   For the optimization and background studies we use only events that fall within the ``analysis window'' defined by $\lowAWmes < \mes < \highAWmes\gevcc$ and $|\DeltaE| < \highAWdele\gev$; this defines the range of the histograms in Fig.~\ref{fig:result}.  A perfectly reconstructed \mydecay\ decay should have $\DeltaE = 0$ and $\mes = m_{B}$.  Therefore we demand that \mydecay\ candidates fall within the ``signal region'' in the \DeltaE\ vs. \mes\ plane defined by $\LSBmes < \mes < \HSBmes\gevcc$ and $\LSBdele < \DeltaE < \HSBdele\gev$.  In Fig.~\ref{fig:result}, the signal region is indicated by a box.

We reject continuum background using a number of topological variables to distinguish between continuum events, which tend to be highly directional, and \B-decay events, which tend to be spherically symmetric.  We determine the thrust and sphericity axes of the particles not used to reconstruct the \B\ candidate, and demand that the angle $\theta_t$ ($\theta_{sph}$) between the thrust (sphericity) axis of these particles and the thrust (sphericity) axis of the \B\ candidate satisfy $|\costhetathrust| < \costbbzhigh$ ($|\costhetasph| < \costhsphbzhigh$).  We further demand that the polar angle $\theta_B$, the angle between the beam direction and the flight direction of the \B\ candidate in the \epem\ center-of-mass frame, satisfy $|\costhetaB| < \coshelbzhigh.$  This demand accepts or rejects events based on the angular momentum constraints of the decay \upsbb.  Finally, we also tighten the $R_2$ requirement to $R_2 < 0.45$.  Studies both of simulated background and off-peak data indicate that the fraction of continuum events satisfying these criteria is negligible.

We reject background from \B\ decays using \jpsi\ and photon selection criteria.  For the \jpsi\ selection, the invariant mass of the \ellell\ pair of the reconstructed \jpsill\ decay is required to fall close to that of the known \jpsi\ mass \cite{ref:PDG}: $ \mjeelow < m(\epem) < \mjeehigh\gevcc $ for \jpsiee\ candidates and $\mjmumulow <  m(\mumu) < \mjmumuhigh\gevcc $ for \jpsimumu\ candidates.  We require that photon candidates satisfy $LAT < \LAThigh$, where $LAT$ \cite{ref:Drescher} is a shower-shape variable used to distinguish between electromagnetic and hadronic showers.  In addition, we constrain the photon direction to the region $\cos\theta_{\gamma} > \costhetagamlow$.

The main source of photons in \babar\ is the decay of neutral pions, so we apply a veto to reject photons from \pigg\ decays.  We reject events in which the \mydecay\ photon candidate combined with any other photon candidate forms a pair with an invariant mass within 20\mevcc of the neutral pion mass \cite{ref:PDG}. 
The signal efficiency of the optimized selection is estimated from the simulations to be $\eff = 0.102 \pm 0.010$.

\begin{table}
\centering
\caption{The selection criteria.}
\begin{tabular}{ll}
\hline \hline Variable & Requirement \\
\hline \jpsi\ mass & $ \mjeelow < m(\epem) < \mjeehigh\gevcc $ \\
& $\mjmumulow <  m(\mumu) < \mjmumuhigh\gevcc $ \\
photon $LAT$ & $LAT < \LAThigh$ \\
photon angle & $\cos\theta_{\gamma} > \costhetagamlow$ \\
\piz\ veto & reject \ $0.115 < m_{\gamma pair} < 0.155\gevcc$ \\
Fox Wolfram moment & $R_2 < \R2allhigh$ \\
thrust angle & $|\costhetathrust| < \costbbzhigh$ \\
sphericity angle & $|\costhetasph| < \costhsphbzhigh$ \\
\B polar angle & $|\costhetaB| < \coshelbzhigh$ \\
\hline signal region & $\LSBmes < \mes < \HSBmes\gevcc$ \\
 & $\LSBdele < \DeltaE < \HSBdele\gev$ \\
\hline \hline
\end{tabular}
\label{tab:selection}
\end{table}

Of interest in the background studies are the events that pass all of the selection criteria except for the requirement to fall within the signal region (Fig.~\ref{fig:result}c).  Most of this background is concentrated in the low-\DeltaE\ region of the \DeltaE-\mes\ plane.  The asymmetry of the signal region in \DeltaE\ ensures that the majority of these events fall outside of the signal region.  The small fraction of this background in the signal region is due primarily to \Bz\to\jpsi\piz\ decays in which a photon from \pigg\ is misidentified as a \mydecay\ photon.  This usually occurs when the other photon in the reconstruction falls below the 30\mev energy threshold.  There is also background from \Bz\to\jpsi\KL\ decays, due to $\KL\to 3\piz$ decays in the EMC for which the six resulting showers overlap and are incorrectly interpreted as a shower from a single photon.

We estimate the background using a large simulated sample distinct from that used to optimize the selection criteria.  Each event in this sample contains either a $B\to\jpsi\piz$ or a $B\to\jpsi\KL$ decay.  After normalizing to the data luminosity we obtain background estimates of 0.59 in the $B\to\jpsi\piz$ mode and 0.12 in the $B\to\jpsi\KL$ mode, resulting in a total background estimate of $n_b = 0.71 \pm 0.31$ events.  The contributions to the uncertainty are discussed below.

To validate the simulated-background modeling we perform several cross-checks.  We compare background estimates from simulations and from on-peak data, outside the signal region but in the analysis window.  The results are consistent both when the estimates are obtained with all of the selection criteria applied, and when the estimates are obtained with all of the criteria applied except for the pion veto.  In addition, we compare the background estimates from off-peak data and from simulated continuum background in the full analysis window.  In both cases, no events pass the selection criteria.

The relative systematic errors in the signal efficiency and in the background estimate are presented in Table~\ref{tab:systematic}.  For both \eff\ and $n_b$ there is statistical uncertainty in the number of events passing the selection.   The uncertainty in the background estimate also includes uncertainty from the number of \FourS\ in the data set, $N_{\FourS} = (123.3 \pm 1.4) \times 10^{6}$, and the uncertainty in the following branching fractions.  \BR($\Bz\to\jpsi\piz$) and \BR($\Bz\to\jpsi\KL$) are obtained from Ref.~\cite{ref:PDG}.  $\BR(\jpsill) = 0.1181 \pm 0.0020$ is the sum of the \jpsiee\ and \jpsimumu\ branching fractions \cite{ref:PDG} assuming fully correlated uncertainties.  $B[\myupsbzbz] = 0.499 \pm 0.012$ is determined from Ref.~\cite{ref:BB} assuming that the \FourS\ decays 100\% to \BB.

In addition, we correct for differences between simulations and data, and each of these corrections contributes to the systematic uncertainty.   The required corrections for tracking, lepton-identification, and photon-reconstruction efficiencies are derived from independent studies comparing the results from simulations with those from data control samples.  Also, comparison of the \DeltaE\ distribution of $\Bz\to\Kstarz\gamma$ decays in real and simulated samples reveals a difference of about 28\mev in the central value for \DeltaE\ between data and Monte Carlo.  This effect is due to imperfect simulation of photon energy loss in the detector.  \mydecay\ is topologically similar to $\Bz\to\Kstarz\gamma$ but has a lower photon energy, so we apply a correction of ($22 \pm 10\mev$) to \DeltaE\ in the simulated samples.  As shown in Table~\ref{tab:systematic}, this \DeltaE correction leads to the largest systematic error in both the efficiency and the background calculation.

\begin{table}
\caption{Summary of relative systematic uncertainties on the signal efficiency $\eff$ and background estimate $n_b$.}
\begin{tabular}{lcc}
\hline \hline & \multicolumn{2}{c}{Uncertainty (\%)} \\
\cline{2-3} Source & \eff & $n_b$ \\
\hline \DeltaE\ correction & 7.6 & 33 \\
Tracking  & 4.0 & 3.4 \\
Lepton ID & 3.5 & 3.2 \\
Neutral ID & 2.5 & 2.1 \\
Statistics (simulated samples) & 1.5 & 24 \\
\BR($\Bz\to\jpsi\piz$), \BR($\Bz\to\jpsi\KL) $ & N/A & 15 \\
\ \ \ \BR(\jpsill), and \BR[\myupsbzbz]  & & \\
\hline Total & 9.8 & 44 \\
\hline \hline
\end{tabular}
\label{tab:systematic}
\end{table}

No events in the signal region satisfy the final selection criteria (Fig.~\ref{fig:result}a).  The probability of observing 0 events when expecting a background of 0.71 events is 49\%.  In the analysis window we observe 10 events in data, consistent at the 8\% level with the expected background of 5.7+/-1.0 events.

We determine the upper limit on the branching fraction \BR(\mydecay) by performing a Bayesian analysis with a uniform prior above zero.  We define the likelihood for \BR(\mydecay) as the probability that exactly zero events pass the selection, given that the mean expected number of observed events is
\begin{equation}
\mu_\nu = n_b + N_{\Bz} \: \eff\ \: \BR(\jpsill) \: \BR(\mydecay),
\end{equation}
where $N_{\Bz} = 2 \: N_{\Upsilon(4S)} \: B[\myupsbzbz]$ is the number of \Bz\ mesons in the data set.  The analysis takes into account the uncertainties in \eff\ and $n_b$.  The 90\% confidence level upper limit, defined as the branching fraction value that separates the lower 90\% of the area under the likelihood function curve from the upper 10\%, is $\BR(\mydecay) < \newuplim$.  This limit is dominated by statistical errors; in the absence of systematic errors, it would improve by less than $0.1 \times 10^{-6}$.

% Standard acknowledgments paragraph; must always be included.
We are grateful for the 
extraordinary contributions of our \pep2\ colleagues in
achieving the excellent luminosity and machine conditions
that have made this work possible.
The success of this project also relies critically on the 
expertise and dedication of the computing organizations that 
support \babar.
The collaborating institutions wish to thank 
SLAC for its support and the kind hospitality extended to them. 
This work is supported by the
US Department of Energy
and National Science Foundation, the
Natural Sciences and Engineering Research Council (Canada),
Institute of High Energy Physics (China), the
Commissariat \`a l'Energie Atomique and
Institut National de Physique Nucl\'eaire et de Physique des Particules
(France), the
Bundesministerium f\"ur Bildung und Forschung and
Deutsche Forschungsgemeinschaft
(Germany), the
Istituto Nazionale di Fisica Nucleare (Italy),
the Foundation for Fundamental Research on Matter (The Netherlands),
the Research Council of Norway, the
Ministry of Science and Technology of the Russian Federation, and the
Particle Physics and Astronomy Research Council (United Kingdom). 
Individuals have received support from 
CONACyT (Mexico),
the A. P. Sloan Foundation, 
the Research Corporation,
and the Alexander von Humboldt Foundation.

\end{document}